\documentclass[runningheads]{llncs}
\usepackage[utf8]{inputenc}

\usepackage[table,dvipsnames]{xcolor}

\usepackage{tikz,times}
\usepackage{float}
\usepackage{hyperref}
\usetikzlibrary{arrows,shapes,topaths,calc,plotmarks,patterns, decorations.pathreplacing, chains, fit, calc, shapes,decorations.markings,chains}

\usepackage{adjustbox}

\usepackage{amsmath}
\usepackage{amssymb}
\usepackage{placeins}
\usepackage{comment}
\usepackage{subfig}
\newcommand\m{\mathcal}

\newcommand\setn{\{1,\dots,n\}}

\newcommand\contn{(\m S_1,\dots,\m S_n,\m R)}
\newcommand\concn{(X_1,\dots,X_n)}

\newcommand\x{$\times$}

\begin{document}

\title{Some Notes on Polyadic Concept Analysis}
\author{Alexandre Bazin\inst{1}, Giacomo Kahn\inst{2}, Camille Noûs\inst{3}}

\institute{LIRMM, CNRS, Université de Montpellier, FRANCE\\
\email{alexandre.bazin@u-montpellier.fr}\\
\and
Univ Lyon, Univ Lyon 2, Université Claude Bernard Lyon 1, Université Jean Monnet Saint-Etienne, INSA Lyon, DISP-UR4570, 69500 Bron, FRANCE\\
\email{giacomo.kahn@univ-lyon2.fr}\\
\and
Cogitamus Laboratory, FRANCE\\
\url{cogitamus.fr/camilleen.html}}

\maketitle

\begin{abstract}

Despite the popularity of Formal Concept Analysis (FCA) as a mathematical framework for data analysis, some of its extensions are still considered arcane. Polyadic Concept Analysis (PCA) is one of the most promising yet understudied of these extensions. This formalism offers many interesting open questions but is hindered in its dissemination by complex notations and a lack of agreed-upon basic definitions. In this paper, we discuss in a mostly informal way the fundamental differences between FCA and PCA in the relation between contexts, conceptual structures, and rules. We identify open questions, present partial results on the maximal size of concept $n$-lattices and suggest new research directions.

\end{abstract}

\section{Unde Venis Et Quo Vadis?}

Formal Concept Analysis (FCA~\cite{ganter2012formal}) is a formalism that establishes a connection between classical binary data (crosstables) and the structure of concepts and rules that can be found in said data. It is very powerful, if underutilised, as it offers well-studied mathematical structures to be exploited by algorithms.

As crosstables are a rather limiting way of representing data, various extensions of the formalism have been proposed to deal with more complex data, such as Pattern Structures~\cite{ganter2001pattern}, Relational Concept Analysis~\cite{rouane2013relational}, fuzzy FCA~\cite{poelmans2014fuzzy} or graph FCA~\cite{ferre2020graph}.
Just like FCA, they are based on lattice theory~\cite{birkhoff1940lattice}.

Triadic Concept Analysis~\cite{lehmann1995triadic} and Polyadic Concept Analysis (PCA)~\cite{voutsadakis2002polyadic} aim to extend FCA to data in the form of $n$-ary relations (\textit{i.e.} multidimensional crosstables) and have the peculiarity of involving $n$-lattices instead of lattices. 
Such structures are considerably less known and studied, and results that would be considered basic in lattice theory are missing.
While some would consider this an opportunity to pioneer a whole new field, the heavy notations and lack of clear definitions and impressive applicative results repels researchers.
The opportunities are however numerous as multidimensional data is now ubiquitous: RDF datasets, folksonomies, pharmacogenomical knowledge are all inherently at least triadic and transforming them to fit dyadic crosstables only results in lost information. As such data are also understudied in data mining in general, this is a unique occasion to position FCA and its extensions as a leading formalism. Some work has already been done in this direction in the field of multidimensional association rule mining~\cite{missaoui2022computing,bazin2022condensed}.

In this paper, we aim at promoting Polyadic Concept Analysis by discussing the differences with FCA introduced by $n$-lattices, suggesting future research directions, and presenting recent results and open questions on the size of $n$-lattices. Section~\ref{sec:preliminaries} contains definitions, including a proposition for a stable definition of implications. Section~\ref{sec:pcaAndTheFallOfGeneralisation} discusses the impact of the loss of duality between orders in $n$-lattices on the relation between FCA structures, as well as a suggestion for a multidimensional generalisation of Boolean lattices. Section~\ref{sec:countingStuff} presents recent results on the maximal size of concept $n$-lattices found in cubic contexts.

\section{The Basics of Boxes\label{sec:preliminaries}}

We assume that the reader is at least somewhat familiar with FCA. If this is not the case, we refer the reader to the usual book~\cite{ganter2012formal}.
Polyadic concept analysis is a natural extension of FCA in which the underlying relation is $n$-ary. It was first introduced in its triadic form by Lehman and Wille~\cite{lehmann1995triadic}, and then generalised to the $n$-dimensional case by Voutsadakis~\cite{voutsadakis2002polyadic}.
In this setting, an $n$-context is an $(n+1)$-tuple $\contn$ where the $\m S_i$ are sets called dimensions and $\m R\subseteq \prod_{i=1}^{n} \m S_i$ is an $n$-ary relation between them. We shall call \emph{objects} the elements of the first dimension $\m S_1$ while the $(n-1)$-tuples $(s_2,\dots,s_n)$ such that $(s_1,s_2,\dots,s_n)\in \m R$ form the \emph{description} of the object $s_1$. An example of a triadic context is depicted in Fig.~\ref{fig:ncont}.

\begin{figure}[htb]
\centering
\begin{tabular}{c|ccc||ccc||ccc}
     & ~$a$~ & ~$b$~ & ~$c$~ & ~$a$~ & ~$b$~ & ~$c$~ & ~$a$~ & ~$b$~ & ~$c$~ \\
     \hline
    ~$1$~ & \x & & & \x & & \x & & & \x \\
    ~$2$~ & \x & & & \x & & & & & \\
    ~$3$~ & \x & \x & & & \x & & & \x & \x \\
     \hline
     \multicolumn{1}{c|}{} & \multicolumn{3}{c||}{$\alpha$} & \multicolumn{3}{c||}{$\beta$} &
     \multicolumn{3}{c}{$\gamma$} 
\end{tabular}
\caption{\label{fig:ncont}A triadic context $(\{\alpha,\beta\},\{1,2,3\},\{a,b,c\},\m R)$.}
\end{figure}

An $n$-concept is then a maximal $n$-dimensional box full of crosses, \textit{i.e.} an $n$-tuple $\concn$ such that $\prod_{i=1}^{n} X_i\subseteq \m R$ and no $X_i$ is such that 
$$\prod_{j=1}^{i-1}\m S_j\times X_i\cup \{x\}\times \prod_{k=i+1}^{n}\m S_k\subseteq \m R$$ with $x\not\in X_i$.
For instance, $(\{\alpha,\beta\},\{1,2\},\{a\})$ is an $n$-concept in Fig.~\ref{fig:ncont}. The set $\mathcal C$ of all $n$-concepts in an $n$-context together with the $n$ quasi-orders induced by the inclusion relation on their $n$ components form an $n$-lattice $\mathcal L = (\mathcal C,\lesssim_1,\dots,\lesssim_n)$.
Note that an $n$-lattice is an $n$-ordered set, \textit{i.e.} it respects:
\begin{itemize}
    \item if $\forall i\in \{1,\dots,n\}\setminus \{j\}, A\lesssim_i B$ then $A\gtrsim_j B$ (antiordinal dependency) and
    \item if $\forall i\in\setn$, $A\sim_i B$, then $A = B$ (uniqueness condition).
\end{itemize}

Different definitions of implications in triadic and polyadic contexts have been proposed through the years such as ``Biedermann's implications''~\cite{biedermann1997triadic}, attribute$\times$condition, conditional attribute or attributional condition implications~\cite{ganter2004implications}.
In~\cite{bazin2020implication}, it was proposed to consider all the implications that hold in dyadic contexts resulting from combinations of two transformations of an $n$-context $\m C$ (see Fig.~\ref{fig:transfo}):
\begin{itemize}
    \item $\m C^{(A,B)} = (\prod A,\prod B,\mathcal R^{(A,B)})$, where $A$ and $B$ form a bipartition of the set of all dimensions and $((s_{a_1},\dots,s_{a_k}),(s_{b_1},\dots,s_{b_l}))\in \m R^{(A,B)}$ iff $(s_1,\dots,s_n)\in \m R$ with $\{a_1,\dots,a_k\}\cup \{b_1,\dots,b_l\} = \{1,\dots,n\}$ 
    \item $\m C_{D} = (\m S_1,\dots,\m S_{d-1},\m S_{d+1},\dots,\m S_n,\m R_D)$, where $D\subseteq \m S_d$ and $$(s_1,\dots,s_{d-1},s_{d+1},\dots,s_n)\in \m R_D \text{ iff }\forall x\in D, (s_1,\dots,s_{d-1},x,s_{d+1},\dots,s_n).$$
\end{itemize}

\begin{figure}[htb]
\centering
\begin{tabular}{c|ccccccccc}
     & ~$(1,a)$~ & ~$(1,b)$~ & ~$(1,c)$~ & ~$(2,a)$~ & ~$(2,b)$~ & ~$(2,c)$~ & ~$(3,a)$~ & ~$(3,b)$~ & ~$(3,c)$~ \\
     \hline
    ~$\alpha$~ & \x & & & \x & & & \x & \x & \\
    ~$\beta$~ & \x & & \x & \x & & & & \x & \\
    ~$\gamma$~ & & & \x & & & & & \x & \x 
     
\end{tabular}

\bigskip
\begin{tabular}{cc}
\parbox{0.4\textwidth}{\hspace{0.18\textwidth}
    \begin{tabular}{c|ccc}
     & ~$1$~ & ~$2$~ & ~$3$~ \\
     \hline
    ~$\alpha$~ & \x & \x & \x \\
    ~$\beta$~ & \x & \x & \\
    ~$\gamma$~ & & & \\
    \end{tabular}
    }
     &  
\parbox{0.4\textwidth}{\hspace{0.1\textwidth}
     \begin{tabular}{c|ccc}
     & ~$a$~ & ~$b$~ & ~$c$~ \\
     \hline
    ~$\alpha$~ & \x & & \\
    ~$\beta$~ & & & \\
    ~$\gamma$~ & & & \x \\ 
     \end{tabular}
     }
\end{tabular}

\caption{\label{fig:transfo}Transformations $\m C^{(\{greek\},\{numbers,latin\})}$ (top), $\m C_{\{a\}}$ (bottom left) and $\m C_{\{1,3\}}$ (bottom right) of Fig.~\ref{fig:ncont}'s triadic context $\m C$.}
\end{figure}

Such implications contain all previously defined types of implications. For instance, in Fig.~\ref{fig:ncont}'s triadic context $\m C$ and if we assume that $numbers = \{1,2,3\}$ is the set of attributes and $latin = \{a,b,c\}$ is the set of conditions, 
\begin{itemize}
\item conditional attribute implications $A_2\xrightarrow{C} A_2$ are the dyadic implications in  the $2$-context $\m C_{C}^{(\{greek\},\{numbers\})}$, \textit{e.g.} $\{3\}\xrightarrow{\{a\}} \{1,2\}$, 
\item attributional condition implications $C_1\xrightarrow{A} C_2$ are the dyadic implications in the $2$-context  $\m C_{A}^{(\{greek\},\{latin\})}$, \textit{e.g.} $\emptyset\xrightarrow{\{3\}} \{b\}$,
\item and attribute$\times$condition implications $A\rightarrow B$ are simply the dyadic implications in  the $2$-context $\m C^{(\{greek\},\{numbers,latin\})}$, \textit{e.g.} $\{(1,a)\}\rightarrow \{(3,b)\}$. 
\end{itemize}

Note that possible implications include those in $\m C^{(\{greek,numbers\},\{latin\})}$ whose support is in a Cartesian product of dimensions.
Such rules (under the more general umbrella of association rules) are already under consideration in the data mining community~\cite{nguyen2011multidimensional,9720169}.
Additionally, if one is only interested in implications that do not contain the first dimension -- the objects -- these implications are all derivable from the implications of $\m C^{(\{\m S_1\},\{S_2,\dots,S_n\})}$ through the application of Armstrong's axioms plus two other axioms, as discussed in~\cite{bazin2020implication}. This means that one only has to reason on a single type of implication.

Just as in the dyadic case, the set of all the implications that do not involve the first (object) dimension and that hold in an $n$-context (or any its implication bases) can be used to reconstruct the set of $n$-concepts restricted to their last $n-1$ components~\cite{bazin2020implication}.

\medskip

We conclude this first section with a small digression about graphical representation.
In two dimensions, both partial orders of the concept lattice $(\m T(\m C), \subseteq_1, \subseteq_2)$ have the decency of being dual, and thus one can be omitted, allowing a concept lattice to be graphically represented by a Hasse diagram.
Starting from three dimensions, concepts are ordered differently.
In $n$ dimensions, there are classes of equivalent concepts with the same $i$th component, and those concepts can then be differentiated in the $n-1$ other quasi-orders.
A graphical representation that clearly shows the equivalence classes and the quasi-order relations is -- at best -- hard to attain.
This is a first open question offered by PCA.

In three dimensions, a tentative graphical representation exists~\cite[Section 3]{lehmann1995triadic}, that combines a geometric representation of equivalence classes together with Hasse-like representation for each quasi-order.
The few graphical representations of $4$-lattices, for example in~\cite[figures 1 and 2]{voutsadakis2002polyadic} can only make one hope for another type of graphical representation, possibly \emph{via} Virtual Reality, as is already proposed for FCA~\cite{sacarea2020improving,breckner2022improving}.

\section{Loss of Duality and the FCA Trinity\label{sec:pcaAndTheFallOfGeneralisation}}

\subsection{Structural Equivalences}

In the dyadic case, the formal context, the concept lattice and the set of implications are equivalent in the sense that they can be computed from one another. This is the reason why the FCA formalism is so useful for data analysis. In the multidimensional case, this equivalence softens, which introduces new challenges, constraints and open questions. Let us consider the two triadic contexts depicted in Fig.~\ref{fig:deuxcont} in which the Greek letters form the first dimension (the objects), and their triadic concepts.

\begin{figure}[htb]
\centering

\begin{tabular}{c | c}

\parbox{0.45\textwidth}{\centering
\begin{tabular}{c|ccc||ccc}
 & ~$a$~ & ~$b$~ & ~$c$~ & ~$a$~ & ~$b$~ & ~$c$~  \\
 \hline
~$1$~ & $\times$ & $\times$ & $\times$ & $\times$ & &  \\
~$2$~ & $\times$ &  & &  & $\times$ &   \\
~$3$~ & $\times$ & &  & &  &  \\
\hline
\multicolumn{1}{c|}{} & \multicolumn{3}{c||}{$\alpha$} & \multicolumn{3}{c}{$\beta$} \\
\end{tabular}
}

&

\parbox{0.45\textwidth}{\centering
\begin{tabular}{c|ccc||ccc||ccc}
 & ~$a$~ & ~$b$~ & ~$c$~ & ~$a$~ & ~$b$~ & ~$c$~ & ~$a$~ & ~$b$~ & ~$c$~ \\
 \hline
~$1$~ & $\times$ & $\times$ & $\times$ & $\times$ & & & $\times$ & & \\
~$2$~ & & & & $\times$ &  & &  & $\times$ & \\
~$3$~ & & & & $\times$ & & & & & \\
\hline
\multicolumn{1}{c|}{} & \multicolumn{3}{c||}{$\alpha$} & \multicolumn{3}{c||}{$\beta$} & \multicolumn{3}{c}{$\gamma$}\\
\end{tabular}
} \\

\multicolumn{2}{c}{}\\
\multicolumn{2}{c}{}\\

\parbox{0.45\textwidth}{\centering
($\emptyset$,$\{1,2,3\}$,$\{a,b,c\}$)

($\{\alpha,\beta\}$,$\emptyset$,$\{a,b,c\}$)

($\{\alpha,\beta\}$,$\{1,2,3\}$,$\emptyset$)

($\{\alpha\}$,$\{1\}$,$\{a,b,c\}$)

($\{\alpha\}$,$\{1,2,3\}$,$\{a\}$)

($\{\alpha,\beta\}$,$\{1\}$,$\{a\}$)

($\{\beta\}$,$\{2\}$,$\{b\}$)
}

&

\parbox{0.45\textwidth}{\centering
($\emptyset$,$\{1,2,3\}$,$\{a,b,c\}$)

($\{\alpha,\beta,\gamma\}$,$\emptyset$,$\{a,b,c\}$)

($\{\alpha,\beta,\gamma\}$,$\{1,2,3\}$,$\emptyset$)

($\{\alpha\}$,$\{1\}$,$\{a,b,c\}$)

($\{\beta\}$,$\{1,2,3\}$,$\{a\}$)

($\{\alpha,\beta,\gamma\}$,$\{1\}$,$\{a\}$)

($\{\gamma\}$,$\{2\}$,$\{b\}$)
}
\end{tabular}

\caption{\label{fig:deuxcont}Two formal contexts $(\{\alpha,\beta,\gamma\},\{1,2,3\},\{a,b,c\},\mathcal R_1)$ (left) and $(\{\alpha,\beta\},\{1,2,3\},$ $\{a,b,c\},\mathcal R_2)$ (right) and their associated $3$-concepts.}
\end{figure}

We observe that the concepts differ only on their first components. Hence, the second and third quasi-orders of both triadic concept lattices are isomorphic while the first quasi-orders are not. This is a most significant change from the bidimensional case: the knowledge of $n-1$ quasi-orders is not enough to know the last one. Let us call the last $n-1$ components of a concept its \emph{feature} and the first component its \emph{extent}, as usual. Then, given a set of concepts known only by their features, there are multiple non-isomorphic ways objects can belong to the extents. This clashes with the usual notions of subsumption.

Let us say that two concept $n$-lattices are equivalent if and only if their $k$th quasi-orders are isomorphic for all $k\in \{2,\dots,n\}$. We denote by $[\mathcal L]$ the equivalence class of the concept $n$-lattice $\mathcal L$. As mentioned in the previous section, implications in $n$-contexts can be used to construct (exactly) the features of the associated $n$-concepts. Hence, all the implication bases of all the $n$-contexts of the lattices in $[\mathcal L]$ allow for the construction of the same features. These implications can, however, differ from $n$-context to $n$-context. For instance, the implication $\{(2,a)\}\rightarrow \{(1,b)\}$ holds in Fig.~\ref{fig:deuxcont}'s first triadic context (as both crosses appear together in the description of the object $\alpha$) but not in the second. From this, we deduce that some implications have no influence on the construction of the features. We thus propose to identify two types of implications:
\begin{itemize}
    \item \emph{Structural implications} that are used to construct the features of all the $n$-concepts of an $n$-context
    \item \emph{Contextual implications} that are not structural but still hold in an $n$-context
\end{itemize}

Structural implications carry information about the features of $n$-concepts while contextual implications carry information about the distribution of the objects in the $n$-concepts. Thus, structural implications are common to all the $n$-contexts in an equivalence class $[\mathcal L]$ while contextual implications are not. In Fig.~\ref{fig:deuxcont}'s contexts, the implications $\{(1,b),(1,c)\}\rightarrow \{(1,a)\}$ and $\{(2,a)\}\rightarrow \{(3,a)\}$ are structural while $\{(2,b)\}\rightarrow \{(1,a)\}$ is contextual in both contexts and $\{(1,b),(2,a)\}\rightarrow \{(3,a),(1,c)\}$ is contextual in the first context only. 
In~\cite{bazin2020implication} is explained in an overly formal way, for which the author is very sorry, that constructing the features of all the $n$-concepts of an $n$-context only requires implications between ``boxes'', \emph{i.e.} implications of the form $\prod_{k = 2}^n X_k\rightarrow \prod_{k = 2}^n Y_k$ with $X_k\subseteq Y_k\subseteq \mathcal S_k$. The structural implications are all the implications entailed by these implications between ``boxes''.

In an equivalence class $[\m L]$, there is one $n$-lattice/$n$-context that seems to be of particular interest as it minimises the number of contextual implications with a non-empty support, which could have some use in data mining. It is fairly easy to construct this $n$-context: if two $n$-concept features are such that their intersection is not the feature of an $n$-concept, then the two features must appear in the descriptions of the same objects. This defines equivalence classes of features. Then, each such equivalence class of features is used to describe a different object. Fig.~\ref{fig:structurality} illustrates this. The rectangles $(\{1\},\{a,b\})$ and $(\{1,2\},\{a\})$ are features of triadic concepts while $(\{1\},\{a\})$, their intersection, is not so both rectangles have to appear in the description of the same object. The intersection $(\{3\},\{c\})$ of the rectangles $(\{2,3\},\{c\})$ and $(\{3\},\{b,c\})$ is the feature of the triadic concept $(\{\alpha,\beta\},\{3\},\{c\})$ so the three rectangles are put in the descriptions of different objects. 

\begin{figure}[htb]

\begin{tabular}{clc}
\parbox{0.4\textwidth}{\centering
\begin{tabular}{c|ccc||ccc}
 & ~$a$~ & ~$b$~ & ~$c$~ & ~$a$~ & ~$b$~ & ~$c$~  \\
 \hline
~$1$~ & $\times$ & $\times$ &  &  & &  \\
~$2$~ & $\times$ &  & &  &  & $\times$ \\
~$3$~ &  & & $\times$ & & $\times$ & $\times$ \\
\hline
\multicolumn{1}{c|}{} & \multicolumn{3}{c||}{$\alpha$} & \multicolumn{3}{c}{$\beta$} \\
\end{tabular}
}
& $\Longrightarrow$ &
\parbox{0.5\textwidth}{\centering
\begin{tabular}{c|ccc||ccc||ccc||ccc}
 & ~$a$~ & ~$b$~ & ~$c$~ & ~$a$~ & ~$b$~ & ~$c$~ & ~$a$~ & ~$b$~ & ~$c$~ & ~$a$~ & ~$b$~ & ~$c$~ \\
 \hline
~$1$~ & $\times$ & $\times$ &  &  & & &  & & &  & &  \\
~$2$~ & $\times$ &  & &  &  &  &  & & $\times$ &  & & \\
~$3$~ &  & & & & $\times$ & $\times$ &  & & $\times$ &  & & $\times$ \\
\hline
\multicolumn{1}{c|}{} & \multicolumn{3}{c||}{$\alpha$} & \multicolumn{3}{c||}{$\beta$} & \multicolumn{3}{c||}{$\gamma$} & \multicolumn{3}{c}{$\delta$} \\
\end{tabular}
}

\end{tabular}
\caption{\label{fig:structurality}Transformation of a triadic context in a way that minimises the number of contextual implications with a non-empty support without changing the equivalence class of the associated concept lattice.}
\end{figure}

\subsection{A Generalisation of Boolean Concept Lattices\label{sec:boolCL}}

We believe that studying such equivalence classes of concept $n$-lattices, instead of individual $n$-context/$n$-lattice pairs, is the way to go. We suggest to start with classes that seem to be particularly interesting as they generalise Boolean concept lattices. Let $\m B_{j_1,\dots,j_{n-1}}$ be the class of concept $n$-lattices such that every possible feature on $n-1$ dimensions of size $j_1,\dots,j_{n-1}$ appear in an $n$-concept. For instance, Fig.~\ref{fig:extrem} depicts the $3$-context and $3$-concepts of a member of $\m B_{3,3}$. All rectangles in a $3\times3$ table appear as the feature of a $3$-concept (the empty rectangle appears twice as $(\{1,2,3\},\emptyset)$ and $(\emptyset,\{a,b,c\})$. 

\begin{figure}[htb]
\centering
\begin{tabular}{c|ccc||ccc||ccc||ccc||ccc||ccc}
     & ~$a$~ & ~$b$~ & ~$c$~ & ~$a$~ & ~$b$~ & ~$c$~ & ~$a$~ & ~$b$~ & ~$c$~ & ~$a$~ & ~$b$~ & ~$c$~ & ~$a$~ & ~$b$~ & ~$c$~ & ~$a$~ & ~$b$~ & ~$c$~  \\
     \hline
    ~$1$~ & \x & \x & & \x & & \x & & \x & \x & \x & \x & \x & \x & \x & \x & & &  \\
    ~$2$~ & \x & \x & & \x & & \x & & \x & \x & \x & \x & \x & & & & \x & \x & \x  \\
    ~$3$~ & \x & \x & & \x & & \x & & \x & \x & & & & \x & \x & \x & \x & \x & \x  \\
    \hline
    \multicolumn{1}{c|}{} & \multicolumn{3}{c||}{$\alpha$} & \multicolumn{3}{c||}{$\beta$} & \multicolumn{3}{c||}{$\gamma$} & \multicolumn{3}{c||}{$\delta$} &
    \multicolumn{3}{c||}{$\epsilon$} &
    \multicolumn{3}{c}{$\zeta$} 
\end{tabular}
%
\adjustbox{max width = \textwidth}{
\begin{tabular}{rrrr}
\multicolumn{2}{c}{$(\{\alpha,\beta,\gamma,\delta,\epsilon,\zeta\},\emptyset,\{a,b,c\})$} &
\multicolumn{2}{c}{$(\{\alpha,\beta,\gamma,\delta,\epsilon,\zeta\},\{1,2,3\},\emptyset)$} \\ 
$(\{\alpha,\beta,\delta,\epsilon\},\{1\},\{a\})$ &
$(\{\alpha,\gamma,\delta,\epsilon\},\{1\},\{b\})$ &
$(\{\beta,\gamma,\delta,\epsilon\},\{1\},\{c\})$ &
$(\{\alpha,\delta,\epsilon\},\{1\},\{a,b\})$  \\
$(\{\beta,\delta,\epsilon\},\{1\},\{a,c\})$ &
$(\{\gamma,\delta,\epsilon\},\{1\},\{b,c\})$ &
$(\{\delta,\epsilon\},\{1\},\{a,b,c\})$ &
$(\{\alpha,\beta,\delta,\zeta\},\{2\},\{a\})$ \\
$(\{\alpha,\gamma,\delta,\zeta\},\{2\},\{b\})$ &
$(\{\beta,\gamma,\delta,\zeta\},\{2\},\{c\})$ &
$(\{\alpha,\delta,\zeta\},\{2\},\{a,b\})$ &
$(\{\beta,\delta,\zeta\},\{2\},\{a,c\})$ \\
$(\{\gamma,\delta,\zeta\},\{2\},\{b,c\})$ &
$(\{\delta,\zeta\},\{2\},\{a,b,c\})$ &
$(\{\alpha,\beta,\epsilon,\zeta\},\{3\},\{a\})$ &
$(\{\alpha,\gamma,\epsilon,\zeta\},\{3\},\{b\})$ \\
$(\{\beta,\gamma,\epsilon,\zeta\},\{3\},\{c\})$ &
$(\{\alpha,\epsilon,\zeta\},\{3\},\{a,b\})$ &
$(\{\beta,\epsilon,\zeta\},\{3\},\{a,c\})$ &
$(\{\gamma,\epsilon,\zeta\},\{3\},\{b,c\})$ \\
$(\{\epsilon,\zeta\},\{3\},\{a,b,c\})$ &
$(\{\alpha,\beta,\delta\},\{1,2\},\{a\})$ &
$(\{\alpha,\gamma,\delta\},\{1,2\},\{b\})$ & 
$(\{\beta,\gamma,\delta\},\{1,2\},\{c\})$ \\
$(\{\alpha,\delta\},\{1,2\},\{a,b\})$ &
$(\{\beta,\delta\},\{1,2\},\{a,c\})$ &
$(\{\gamma,\delta\},\{1,2\},\{b,c\})$ &
$(\{\delta\},\{1,2\},\{a,b,c\})$ \\
$(\{\alpha,\beta,\epsilon\},\{1,3\},\{a\})$ &
$(\{\alpha,\gamma,\epsilon\},\{1,3\},\{b\})$ &
$(\{\beta,\gamma,\epsilon\},\{1,3\},\{c\})$ &
$(\{\alpha,\epsilon\},\{1,3\},\{a,b\})$ \\
$(\{\beta,\epsilon\},\{1,3\},\{a,c\})$ &
$(\{\gamma,\epsilon\},\{1,3\},\{b,c\})$ &
$(\{\epsilon\},\{1,3\},\{a,b,c\})$ &
$(\{\alpha,\beta,\zeta\},\{2,3\},\{a\})$ \\
$(\{\alpha,\gamma,\zeta\},\{2,3\},\{b\})$ &
$(\{\beta,\gamma,\zeta\},\{2,3\},\{c\})$ &
$(\{\alpha,\zeta\},\{2,3\},\{a,b\})$ &
$(\{\beta,\zeta\},\{2,3\},\{a,c\})$ \\
$(\{\gamma,\zeta\},\{2,3\},\{b,c\})$ &
$(\{\zeta\},\{2,3\},\{a,b,c\})$ &
$(\{\alpha,\beta\},\{1,2,3\},\{a\})$ &
$(\{\alpha,\gamma\},\{1,2,3\},\{b\})$ \\
$(\{\beta,\gamma\},\{1,2,3\},\{c\})$ &
$(\{\alpha\},\{1,2,3\},\{a,b\})$ &
$(\{\beta\},\{1,2,3\},\{a,c\})$ &
$(\{\gamma\},\{1,2,3\},\{b,c\})$ \\
\multicolumn{4}{c}{$(\emptyset,\{1,2,3\},\{a,b,c\})$}
\end{tabular}
}
\caption{\label{fig:extrem}$3$-context and $3$-concepts of a concept $3$-lattice in $\m B_{3,3}$.}
\end{figure}

The $n$-lattices in $\m B_{j_1,\dots,j_{n-1}}$ contain $(\prod_{i=1}^{n-1} 2^{j_i}-1)+n-1$ concepts. They generalise the Boolean concept lattices $\m B_{j_1}$ as they are extremal in the sense that all possible features appear in concepts, they do not contain any structural implications (contextual implications may exist) and one of the corresponding $n$-contexts resembles the contranominal scale as, for each object, an element of a dimension is completely missing (see Fig~\ref{fig:extrem}). As such, it is always possible, in a context in which the number of objects is at least equal to the sum of the sizes of all the other dimensions, to produce such an extremal $n$-lattice. It appears that $\m B_{2,2}$ can exist on $3$ objects (Fig~\ref{fig:b22}) while $\m B_{3,3}$ cannot exist on $5$ (see Fig.~\ref{fig:b33} for another such extremal $3$-context on $6$ objects). It is currently unknown whether $\m B_{2,2}$ is an exception or other extremal $n$-lattices can exist on one less object. This problem of finding ``maximally compact'' $n$-contexts producing specific $n$-lattices seems relevant as it is tied to other potentially interesting problems such as finding the maximal number of $n$-concepts in an $n$-context of a given size or maximising the number of contextual implications.

\begin{figure}[htb]
\centering
\begin{tabular}{c|cc||cc||cc}
 & ~$a$~ & ~$b$~ & ~$a$~ & ~$b$~ & ~$a$~ & ~$b$~ \\
 \hline
~$1$~ & \x & \x & & \x & \x & \\
~$2$~ & \x & & \x & \x & & \x \\
\hline
\multicolumn{1}{c|}{} & \multicolumn{2}{c||}{$\alpha$} & \multicolumn{2}{c||}{$\beta$} & \multicolumn{2}{c}{$\gamma$}
\end{tabular}
\caption{\label{fig:b22}A $3$-context producing a member of $\m B_{2,2}$ with only three objects.}
\end{figure}

\begin{figure}[htb]
    \centering
\begin{tabular}{c|ccc||ccc||ccc||ccc||ccc||ccc}
     & ~$a$~ & ~$b$~ & ~$c$~ & ~$a$~ & ~$b$~ & ~$c$~ & ~$a$~ & ~$b$~ & ~$c$~ & ~$a$~ & ~$b$~ & ~$c$~ & ~$a$~ & ~$b$~ & ~$c$~ & ~$a$~ & ~$b$~ & ~$c$~  \\
     \hline
    ~$1$~ & \x & \x &\x &    & \x & \x & \x & \x & \x & \x & \x &    & \x &    & \x & \x & \x & \x  \\
    ~$2$~ & \x & \x &\x & \x & \x & \x & \x &    & \x & \x & \x & \x &    & \x & \x & \x & \x &     \\
    ~$3$~ & \x & \x &   & \x & \x & \x & \x & \x & \x &    & \x & \x & \x & \x & \x & \x &    & \x  \\
    \hline
    \multicolumn{1}{c|}{} & \multicolumn{3}{c||}{$\alpha$} & \multicolumn{3}{c||}{$\beta$} & \multicolumn{3}{c||}{$\gamma$} & \multicolumn{3}{c||}{$\delta$} &
    \multicolumn{3}{c||}{$\epsilon$} &
    \multicolumn{3}{c}{$\zeta$} 
\end{tabular}

\caption{\label{fig:b33}Another $3$-context producing a member of $\m B_{3,3}$ with six objects.}
\end{figure}

\FloatBarrier

\section{The Counter-Curse of Dimensionality\label{sec:countingStuff}}

A natural question one can ask themself about such structures as lattices and $n$-lattices is their maximal size. As discussed in the previous subsection, $n$-contexts in which a dimension is as big as the sum of all the others can contain an extremal $n$-lattice, the size of which is known. The question remains open for all other $n$-contexts.

Given an $n$-context $\contn$ with dimensions of equal size $s$, how many $n$-concepts might that context contain, at most?
We call that number $f_n(s)$.
In this section, we discuss this question, first in the bidimensional case where it is already elegantly solved, and then in the general case where it is not.
The state of the art on this question is presented in Fig.~\ref{fig:regionWhereMaxLivesnd}, at the end of the section.

\subsection{All is Clear in 2 Dimensions}

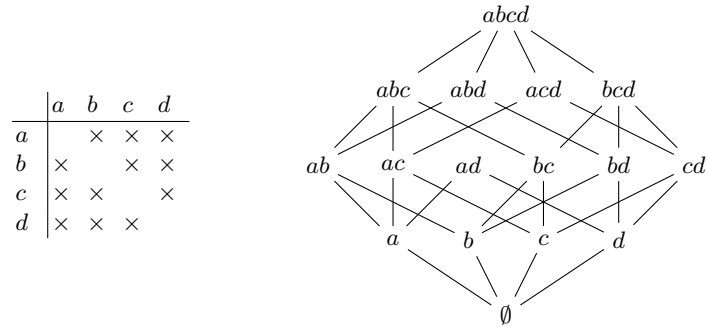
\begin{figure}[htb]
\centering
\begin{minipage}{0.38\textwidth}
\centering
\begin{tabular}{m{0.08\textwidth}|m{0.08\textwidth}m{0.08\textwidth}m{0.08\textwidth}m{0.08\textwidth}}
 & $a$ & $b$ & $c$ & $d$ \\
 \hline
$a$ &  & \x & \x & \x \\
$b$ & \x &  & \x & \x \\
$c$ & \x & \x &  & \x \\
$d$ & \x & \x & \x &  \\
\end{tabular}

\end{minipage}
\begin{minipage}{0.49\textwidth}
\centering
\adjustbox{max width = \textwidth}{
\begin{tikzpicture}
\node (bot) at (0,0) {$\emptyset$};

\node (a) at (-1.5,1) {$a$};
\node (b) at (-0.5,1) {$b$};
\node (c) at (0.5,1) {$c$};
\node (d) at (1.5,1) {$d$};

\node (ab) at (-2.5,2) {$ab$};
\node (ac) at (-1.5,2) {$ac$};
\node (ad) at (-0.5,2) {$ad$};
\node (bc) at (0.5,2) {$bc$};
\node (bd) at (1.5,2) {$bd$};
\node (cd) at (2.5,2) {$cd$};

\node (abc) at (-1.5,3) {$abc$};
\node (abd) at (-0.5,3) {$abd$};
\node (acd) at (0.5,3) {$acd$};
\node (bcd) at (1.5,3) {$bcd$};

\node (top) at (0,4) {$abcd$};

\path
(a) edge (bot)
(a) edge (ab)
(a) edge (ac)
(a) edge (ad)
(b) edge (bot)
(b) edge (ab)
(b) edge (bc)
(b) edge (bd)
(c) edge (bot)
(c) edge (ac)
(c) edge (bc)
(c) edge (cd)
(d) edge (bot)
(d) edge (ad)
(d) edge (bd)
(d) edge (cd)
(ab) edge (abc)
(ab) edge (abd)
(ac) edge (abc)
(ac) edge (acd)
(bc) edge (abc)
(bc) edge (bcd)
(bd) edge (abd)
(bd) edge (bcd)
(cd) edge (acd)
(cd) edge (bcd)
(abc) edge (top)
(abd) edge (top)
(acd) edge (top)
(bcd) edge (top);

\end{tikzpicture}
}
\end{minipage}
\caption{The contranominal scale $\mathbb N_2^c(4)$ and its associated $2$-lattice, a \emph{tesseract} (only the extents are represented).\label{fig:2contraScale}}
\end{figure}

It is well known that $2$-lattices with $2^s$ elements -- powerset lattices, or Boolean lattices -- can be constructed from $2$-contexts of size $s\times s$, ($2$-)contranominal scales.
The $2$-contranominal scale with $s$ objects and $s$ attributes is denoted $\mathbb N_2^c(s)$.
An example of $\mathbb N_2^c(4)$ and its concept lattice is shown in Figure~\ref{fig:2contraScale}.

In~\cite{albano2017concept}, Albano and Chornomaz show that not only do $2$-dimensional contranominal scales $\mathbb N_2^c(s)$ give rise to powerset lattices, but that they are the sole culprit in the large size a concept $2$-lattice can have.
A \emph{large} $2$-lattice has as standard context a context in which lurks a \emph{large} contranominal scale.
In their paper, Albano and Chornomaz quantify exactly what \emph{large} means in both its uses.

\subsection{General Case: What About Dimension $n$?}

First, we can identify a naive upper bound on the maximal number of concept in a $n$-context of size $s\times \dots \times s$.
A $n$-concept is always uniquely described by $(n-1)$ of its components, and as such $( 2^{s}-1)^{n-1}+n-1$ is an upper bound for $f_n(s)$.

Before giving a lower bound for $f_n(s)$, it is necessary to recall some information on the construction of contexts.

The direct sum of two contexts $\mathcal C^1=(\mathcal S_1^1,\dots,S_n^1,\mathcal R^1)$ and $\mathcal C^2=(\mathcal S_1^2,\dots,\mathcal S_n^2,\mathcal R^2)$ where the $S_i^1, S_i^2$ are disjoint is the context $\mathcal C =(\mathcal S_1,\dots,\mathcal S_d,\mathcal R)$ constructed in the following way:

\[\mathcal S_i=\mathcal S_i^1\cup\mathcal S_i^2, \forall i\in\{1,\dots,n\}\]

and
$$\m R=\m R^1\cup \m R^2\cup\{(x_1,\dots,x_d)\mid \exists i, j\in\{1,\dots, n\}, i\neq j \mbox{ such that } x_i\in\mathcal S_i^1, x_j\in\mathcal S_j^2\}.$$

\begin{figure}[htb]
\centering
\begin{minipage}{0.37\textwidth}
\centering
\adjustbox{max width = \textwidth}{
\begin{tikzpicture}[scale=0.6, every node/.style={scale=0.8}]
\pgfmathsetmacro{\cubex}{8}
\pgfmathsetmacro{\cubey}{8}
\node (C12) at (-6, 0.35) {$S_2^1$};
\node (C22) at (-2, 0.35) {$S_2^2$};
\node (C11) at (-8.35, -2) {$S_1^1$};
\node (C21) at (-8.35, -6) {$S_1^2$};
\node (Context1) at (-6, -2) {$\mathcal C^1$};
\node (Context2) at (-2,-6) {$\mathcal C^2$};

\draw[thick] (0,0) -- ++(-\cubex,0) -- ++(0,-\cubey) -- ++(\cubex,0,0) -- cycle;
\coordinate (m1) at (-4,0);
\coordinate (m2) at (-4,-8);
\coordinate (m3) at (0,-4);
\coordinate (m4) at (-8,-4);
\path[] 
	(m1) edge [] (m2)
	(m3) edge [] (m4);

\fill[color=gray!20, pattern=north west lines, very thin](0,0) -- (-4,0) -- (-4,-4) -- (0,-4) -- cycle;
\fill[color=gray!20, pattern=north west lines, very thin](-8,-8) -- (-8,-4) -- (-4,-4) -- (-4,-8) -- cycle;

\fill[color=BurntOrange!69](-5,-2) -- ++ (2.5,0) -- ++ (0,-3) -- ++ (-2.5,0) -- cycle;
\fill[color=OliveGreen!50](-4,-4) -- ++ (-1,0) -- ++ (0,2) -- ++ (1,0) -- cycle;
\fill[color=Bittersweet!50](-4,-4) -- ++ (1.5,0) -- ++ (0,-1) -- ++ (-1.5,0) -- cycle;

\end{tikzpicture}}%
\end{minipage}%
\begin{minipage}{0.5\textwidth}
\centering
\adjustbox{max width = 0.9\textwidth}{
\begin{tikzpicture}[scale=0.6, every node/.style={scale=0.8}]
\pgfmathsetmacro{\cubex}{5}
\pgfmathsetmacro{\cubey}{5}
\pgfmathsetmacro{\cubez}{5}

\fill[color=BurntOrange!69](-3.7,-1,-1.5) -- ++ (0,0,-1.3) -- ++ (1.7,0,0) -- ++ (0,-2.3,0) -- ++ (0,0,1.3) -- ++ (-1.7,0,0) -- cycle;
\fill[color=OliveGreen!50](-3,-2,-2) -- ++ (0,1,0) -- ++ (-0.7,0,0) -- ++ (0,0,0.5) -- ++ (0,-1,0) -- ++ (0.7,0,0) -- cycle;
\fill[color=Bittersweet!50](-3,-2,-2) -- ++ (0,-1.3,0) -- ++ (1,0,0) -- ++ (0,0,-0.8) -- ++ (0,1.3,0) -- ++ (-1,0,0) -- cycle;

\draw[thick] (0,0,0) -- ++(-\cubex,0,0) -- ++(0,-\cubey,0) -- ++(\cubex,0,0) -- cycle;
\draw[thick] (0,0,0) -- ++(0,0,-\cubez) -- ++(0,-\cubey,0) -- ++(0,0,\cubez) -- cycle;
\draw[thick] (0,0,0) -- ++(-\cubex,0,0) -- ++(0,0,-\cubez) -- ++(\cubex,0,0) -- cycle;

\pgfmathsetmacro{\cone}{2}
\pgfmathsetmacro{\ctwo}{3}

\draw[] (-\cubex,0,0) -- ++(\cone,0,0) -- ++(0,-\cone,0) -- ++(-\cone,0,0) -- cycle;
\draw[] (-\cubex,0,0) -- ++(\cone,0,0) -- ++(0,0,-\cone) -- ++(-\cone,0,0) -- cycle;
\draw[dashed] (-\cubex+\cone,0,-\cone) -- ++(0,-\cone,0) -- ++(0,0,\cone);

\draw[dashed] (0,-\cone,-\cubez) -- ++ (-\ctwo,0,0) -- ++(0,0,\ctwo) -- ++(\ctwo,0,0);
\draw[dashed] (-\cubex+\cone,-\cone,-\cone) -- ++(0,-\ctwo,0) -- ++(\ctwo,0,0);
\draw[] (0,-\cone,-\cone) -- ++(0,-\ctwo,0);
\draw[] (0,-\cone,-\cone) -- ++(0,0,-\ctwo);

\node (C1) at (-\cubex+1, -1,-1) {$\mathcal C^1$};
\node (C2) at (-2,-3,-3) {$\mathcal C^2$};

\node (S11) at (-\cubex-1, -1,0) {$S_1^1$};
\coordinate (boutS111) at (-\cubex-0.5, 0, 0);
\coordinate (boutS112) at (-\cubex-0.5, -2, 0);
\draw[<->] (boutS111) -- (boutS112);

\node (S12) at (-\cubex-1, -3,0) {$S_1^2$};
\coordinate (boutS121) at (-\cubex-0.5, -2, 0);
\coordinate (boutS122) at (-\cubex-0.5, -\cubey, 0);
\draw[<->] (boutS121) -- (boutS122);
\node (S21) at (-\cubex, 1,-1) {$S_2^1$};
\coordinate (boutS211) at (-\cubex, 0.5, 0);
\coordinate (boutS212) at (-\cubex, 0.5, -2);
\draw[<->] (boutS211) -- (boutS212);
\node (S21) at (-\cubex, 1,-3) {$S_2^2$};
\coordinate (boutS211) at (-\cubex, 0.5, -2);
\coordinate (boutS212) at (-\cubex, 0.5, -\cubez);
\draw[<->] (boutS211) -- (boutS212);
\node (S31) at (-4, -\cubey-1,0) {$S_3^1$};
\coordinate (boutS311) at (-\cubex, -\cubey-0.5, 0);
\coordinate (boutS322) at (-3, -\cubey-0.5, 0);
\draw[<->] (boutS311) -- (boutS322);
\node (S32) at (-2, -\cubey-1,0) {$S_3^2$};
\coordinate (boutS312) at (-3, -\cubey-0.5, 0);
\coordinate (boutS322) at (0, -\cubey-0.5, 0);
\draw[<->] (boutS312) -- (boutS322);

\end{tikzpicture}}
\end{minipage}
\caption{Illustration of a combination of concepts under a direct sum of two contexts $\m C^1$ and $\m C^2$ in two dimensions (left) and three dimensions (right). Any two concepts of $\m C^1$ and $\m C^2$  -- a maroon and a green one -- can be extended into a new concept of $\m C$ -- through the gray areas --, thus multiplying their number of concepts.\label{fig:directSum}}
\end{figure}
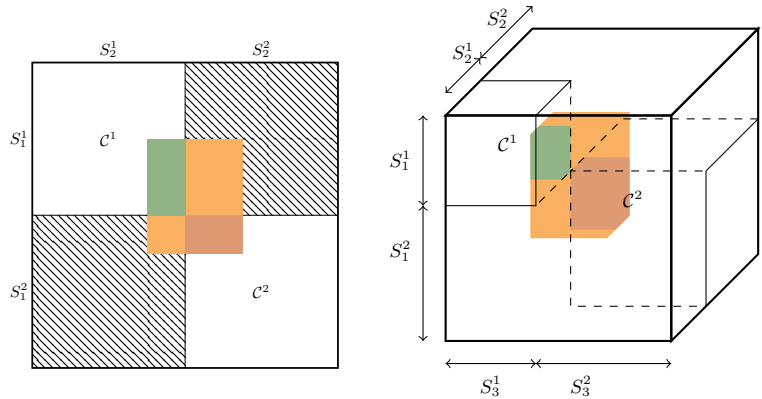

The number of concepts in $\m C$ is then the product of the number of concepts in $\m C^1$ and in $\m C^2$.
This construction is illustrated in two and three dimensions in Fig.~\ref{fig:directSum}.

An empty one-cell $n$-context $\m C_{ell} = (1,\dots, 1, \emptyset)$ gives rise to exactly $n$ $n$-concepts.
Adding together $s$ such empty $n$-contexts gives rise to the $n$-contranominal scale $\mathbb N^c_n(s)$, that is an $n$-context of the form $(\m S,\dots,\m S, R)$ with $R = \prod \m S\setminus \{(x,\dots,x)\mid x\in\m S\}$ that has $n^{|\m S|}$ concepts.
This is a lower bound for $f_n(s)$.

Those two bounds are identical in two dimensions, but are increasingly distant with higher dimensions.

\subsection{In Three Dimensions}

There is no clear answer yet for the value of $f_3(s)$.
The two bounds discussed in the previous section amount to looking for $f_3(s)$ between $3^s$ and $4^s$.

The well known triadic contranominal scale and the associated powerset trilattice were studied by Biedermann~\cite{biedermann1998powerset}.
Powerset trilattices are presented as a triadic generalisation of powerset lattices, can be constructed from $3$-contranominal scales $\mathbb N_3^c(s) = (\m S, \m S, \m S, \neq)$ and have $3^s$ elements.
No explicit claim that this class of $3$-lattices is extremal is made, other than the use of the name powerset~\footnote{In~\cite{biedermann1998powerset}, the name powerset trilattice is justified by some algebraic properties of the powerset trilattice.}.

Although the $3$-contexts responsible for powerset trilattices are good generalisations of contranominal scales, the $3$-lattices themselves are not that good of a generalisation of the extremal $2$-dimensional case, since they are not extremal w.r.t. the size.
This is discussed in Section~\ref{sec:boolCL}.

\begin{figure}[htb]
\centering
\begin{tikzpicture}
\coordinate (c4) at (10,0);
\node[draw, circle, fill = black, inner sep = 1pt] (c3) at (0,0) [label=below:$3^s$]{};
\node[draw, circle, fill = black, inner sep = 1pt] (c4) at (10,0) [label=below:$(2^s - 1)^2 + 2$]{};
\node (c336) at (3.6,0) [label={[xshift=-0.4cm, yshift=-1cm]:$3.36^s$}] {$\vert$};
\node (c339) at (3.9,0) [label={[xshift=0.4cm, yshift=-1cm]:$3.39^s$}] {$\vert$};
\draw (c3) -- (c4);
\end{tikzpicture}
\caption{While either naive bounds would be a satisfying answer for the maximum number of $3$-concepts in a $3$-context, the real answer seems to be in a small corner near  $3.36^s$.\label{fig:regionWhereMaxLives3d}}
\end{figure}
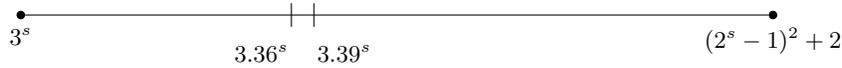

In~\cite{bazin:hal-01847459}, the authors present bounds for the maximum number of $3$-concepts in a $3$-context of size $s\times s\times s$.
Using a small example of a $5\times 5\times 5$ $3$-context that contains more $3$-concepts than a $3$-contranominal scale and the direct sum of contexts, they provide a construction of arbitrarily large contexts with $3.359^s$ concepts.
Using a measure and conquer approach, they also provide an upper bound of $3.384^s$ for $3$-contexts of that size.
The improvement from the naive bounds is represented in Figure~\ref{fig:regionWhereMaxLives3d}.

\subsection{In Four Dimensions}

Now, we extend the work presented in~\cite{bazin:hal-01847459} and provide a lower bound on $f_4(s)$.
The two bounds discussed in the previous section amount, in four dimensions, to looking for $f_4(s)$ between $4^s$ and $8^s$.
Using the same intuition as in~\cite{bazin:hal-01847459}, we build a $3\times3\times3\times3$ $4$-context where the "holes" are a solution to the chess rook problem in $4$ dimensions~\cite{dudeney1970eightrooks}.
This context\footnote{A data file containing the complement of this contexts (its holes) is available at \url{http://giacomo.kahn.science/resources/Crook.dat}} is shown in Table~\ref{tab:4contRook}.

\begin{table}
\centering
\adjustbox{max width = \textwidth}{
\begin{tabular}{c|ccc|ccc|ccc||ccc|ccc|ccc||ccc|ccc|ccc}
 & 1 & 2 & 3 & 1 & 2 & 3 & 1 & 2 & 3 & 1 & 2 & 3 & 1 & 2 & 3 & 1 & 2 & 3 & 1 & 2 & 3 & 1 & 2 & 3 & 1 & 2 & 3 \\
 \hline
1 & \cellcolor{blue!25} & \x & \x & \x & \cellcolor{blue!25} & \x & \x & \x & \cellcolor{blue!25} & \x & \cellcolor{blue!25} & \x & \x & \x & \cellcolor{blue!25} & \cellcolor{blue!25} & \x & \x & \x & \x & \cellcolor{blue!25} & \cellcolor{blue!25} & \x & \x & \x & \cellcolor{blue!25} & \x \\
2 & \x & \cellcolor{blue!25} & \x & \x & \x & \cellcolor{blue!25} & \cellcolor{blue!25} & \x & \x & \x & \x & \cellcolor{blue!25} & \cellcolor{blue!25} & \x & \x & \x & \cellcolor{blue!25} & \x & \cellcolor{blue!25} & \x & \x & \x & \cellcolor{blue!25} & \x & \x & \x &\cellcolor{blue!25} \\
3 & \x & \x & \cellcolor{blue!25} & \cellcolor{blue!25} & \x & \x & \x & \cellcolor{blue!25} & \x & \cellcolor{blue!25} & \x & \x & \x & \cellcolor{blue!25} & \x & \x & \x & \cellcolor{blue!25} & \x & \cellcolor{blue!25} & \x & \x & \x & \cellcolor{blue!25} & \cellcolor{blue!25} & \x & \x \\
\hline
 \multicolumn{1}{c}{} & \multicolumn{3}{c|}{1} & \multicolumn{3}{c|}{2} & \multicolumn{3}{c||}{3} & \multicolumn{3}{c|}{1} & \multicolumn{3}{c|}{2} & \multicolumn{3}{c||}{3} & \multicolumn{3}{c|}{1} & \multicolumn{3}{c|}{2} & \multicolumn{3}{c}{3}\\
 \hline
 \multicolumn{1}{c}{} & \multicolumn{9}{c||}{1} & \multicolumn{9}{c||}{2} & \multicolumn{9}{c}{3}\\
\end{tabular}
}
\caption{This is a $3\times 3\times 3 \times 3$ $4$-context that we call $\m C_{rook}$. Empty cells are coloured blue for readability. This $4$-context has 112 $4$-concepts. This amounts to around $4.82^3$ concepts. This is, just as in the three dimensional case, a solution to a multidimensional chess rook problem.\label{tab:4contRook}}
\end{table}

Using the context $\m C_{rook}$ depicted in Table~\ref{tab:4contRook}, we can build arbitrarily large $4$-contexts that have $c4.82^s$ concepts, with $c$ a constant.

Let $s$ be a integer greater than $3$.
Then, there exists two integers $k$ and $r$ such that $s= 3k+r$, with $r\in[0,2]$.
To build a $4$-context $\m C$ of size $s \times s \times s\times s$, we add $k$ versions of $\m C_{rook}$ with the aforementioned procedure, and then add a $4$-dimensional contranominal scale of size $r$.
The resulting context has $4^r\times 4.82^{s-r} = (\frac{4}{4.82})^r\times 4.82^s$ concepts.
By setting $c = (\frac{4}{4.82})^2$, we have that $f_4(s)\geq c4.82^s$.

\subsection{Summary}

\begin{figure}[htb]
\centering
\input{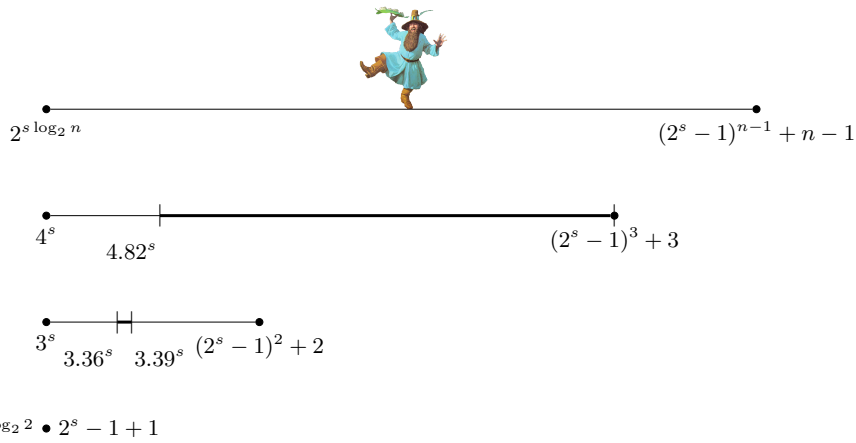}
\caption[Summary of the known bounds for $f_n(s)$]{\emph{"Old Tom Bombadil is a merry fellow! Bright Blue his jacket is, and his boots are yellow!"}. In the general case, we do not really know where the maximum number of $n$-concepts lives. This uncertainty is represented here by a dancing Tom Bombadil. This figures summarizes the known bounds for $f_n(s)$. In two dimensions the bound is well known and reachable, in three dimension the gap is small.\label{fig:regionWhereMaxLivesnd}}
\end{figure}

Figure~\ref{fig:regionWhereMaxLivesnd} summarises the known results related to the size of $n$-lattices, taking into account the contexts from Section~\ref{sec:boolCL} and Section~\ref{sec:countingStuff}.

\bibliographystyle{unsrt}
\bibliography{biblio}

\begin{thebibliography}{10}

\bibitem{ganter2012formal}
Bernhard Ganter and Rudolf Wille.
\newblock {\em {Formal Concept Analysis: Mathematical Foundations}}.
\newblock Springer Science \& Business Media, 2012.

\bibitem{ganter2001pattern}
Bernhard Ganter and Sergei~O. Kuznetsov.
\newblock {Pattern Structures and Their Projections}.
\newblock In {\em International conference on conceptual structures}, pages
  129--142. Springer, 2001.

\bibitem{rouane2013relational}
Mohamed Rouane-Hacene, Marianne Huchard, Amedeo Napoli, and Petko Valtchev.
\newblock {Relational Concept Analysis: Mining Concept Lattices from
  Multi-Relational Data}.
\newblock {\em Annals of Mathematics and Artificial Intelligence},
  67(1):81--108, 2013.

\bibitem{poelmans2014fuzzy}
Jonas Poelmans, Dmitry~I Ignatov, Sergei~O. Kuznetsov, and Guido Dedene.
\newblock {Fuzzy and Rough Formal Concept Analysis: a Survey}.
\newblock {\em International Journal of General Systems}, 43(2):105--134, 2014.

\bibitem{ferre2020graph}
S{\'e}bastien Ferr{\'e} and Peggy Cellier.
\newblock Graph-fca: An extension of formal concept analysis to knowledge
  graphs.
\newblock {\em Discrete applied mathematics}, 273:81--102, 2020.

\bibitem{birkhoff1940lattice}
Garrett Birkhoff.
\newblock {\em {Lattice Theory}}, volume~25.
\newblock American Mathematical Soc., 1940.

\bibitem{lehmann1995triadic}
Fritz Lehmann and Rudolf Wille.
\newblock {A Triadic Approach to Formal Concept Analysis}.
\newblock In {\em International conference on conceptual structures}, pages
  32--43. Springer, 1995.

\bibitem{voutsadakis2002polyadic}
George Voutsadakis.
\newblock {Polyadic Concept Analysis}.
\newblock {\em Order}, 19(3):295--304, 2002.

\bibitem{missaoui2022computing}
Rokia Missaoui, Pedro~HB Ruas, L{\'e}onard Kwuida, Mark~AJ Song, and
  Mohamed~Hamza Ibrahim.
\newblock Computing triadic generators and association rules from triadic
  contexts.
\newblock {\em Annals of Mathematics and Artificial Intelligence}, pages 1--23,
  2022.

\bibitem{bazin2022condensed}
Alexandre Bazin, Nicolas Gros, Aurelie Bertaux, and Christophe Nicolle.
\newblock Condensed representations of association rules in n-ary relations.
\newblock {\em IEEE Transactions on Knowledge and Data Engineering}, 2022.

\bibitem{biedermann1997triadic}
Klaus Biedermann.
\newblock {How Triadic Diagrams Represent Conceptual Structures}.
\newblock In {\em International Conference on Conceptual Structures}, pages
  304--317. Springer, 1997.

\bibitem{ganter2004implications}
Bernhard Ganter and Sergei Obiedkov.
\newblock {Implications in Triadic Formal Contexts}.
\newblock In {\em International Conference on Conceptual Structures}, pages
  186--195. Springer, 2004.

\bibitem{bazin2020implication}
Alexandre Bazin.
\newblock {On Implication Bases in n-Lattices}.
\newblock {\em Discrete Applied Mathematics}, 273:21--29, 2020.

\bibitem{nguyen2011multidimensional}
Kim-Ngan~T Nguyen, Lo{\"\i}c Cerf, Marc Plantevit, and Jean-Fran{\c{c}}ois
  Boulicaut.
\newblock {Multidimensional Association Rules in Boolean Tensors}.
\newblock In {\em Proceedings of the 2011 SIAM International Conference on Data
  Mining}, pages 570--581. SIAM, 2011.

\bibitem{9720169}
Alexandre Bazin, Nicolas Gros, Aurelie Bertaux, and Christophe Nicolle.
\newblock {Condensed Representations of Association Rules in n-ary Relations}.
\newblock {\em IEEE Transactions on Knowledge and Data Engineering}, pages
  1--1, 2022.

\bibitem{sacarea2020improving}
Christian S{\u{a}}c{\u{a}}rea and Raul-Robert Zavaczki.
\newblock {Improving User's Experience in Navigating Concept Lattices: An
  Approach Based on Virtual Reality}.
\newblock 2020.

\bibitem{breckner2022improving}
Brigitte Breckner, Christian S{\u{a}}c{\u{a}}rea, and Raul-Robert Zavaczki.
\newblock {Improving User’s Experience in Exploring Knowledge Structures: A
  Gamifying Approach}.
\newblock {\em Mathematics}, 10(5):709, 2022.

\bibitem{albano2017concept}
Alexandre Albano and Bogdan Chornomaz.
\newblock {Why Concept Lattices Are Large: Extremal Theory for Generators,
  Concepts, and VC-Dimension}.
\newblock {\em International Journal of General Systems}, 46(5):440--457, 2017.

\bibitem{biedermann1998powerset}
Klaus Biedermann.
\newblock {Powerset Trilattices}.
\newblock In {\em International Conference on Conceptual Structures}, pages
  209--221. Springer, 1998.

\bibitem{bazin:hal-01847459}
Alexandre Bazin, Laurent Beaudou, Giacomo Kahn, and Kaveh Khoshkhah.
\newblock {Bounding the Number of Minimal Transversals in Tripartite 3-Uniform
  Hypergraphs}.
\newblock working paper or preprint, January 2022.

\bibitem{dudeney1970eightrooks}
HE~Dudeney.
\newblock {"The Eight Rooks." {\S}295 in Amusements in Mathematics}, 1970.

\end{thebibliography}

\end{document}